\documentclass[aip,preprint]{revtex4-1}
\usepackage{graphicx}
\usepackage{dcolumn}
\usepackage{bm}
\usepackage{color}

\begin{document}
\newcommand{\py}{Ni$_{81}$Fe$_{19}$}


\title{Single-domain shape anisotropy in near-macroscopic Ni$_{80}$Fe$_{20}$ thin-film rectangles}

\author{Yi Li}
\author{Yiran Lu}
\author{and W. E. Bailey}

\affiliation{Materials Science \& Engineering, Dept. of Applied Physics \& Applied Mathematics, Columbia University, New York NY 10027, USA}



\date{\today}

\begin{abstract}

Shape anisotropy provides a simple mechanism to adjust the local bias field in patterned structures. It is well known that for ellipsoidal particles $<1$ $\mu$m in size, a quasi-single domain state can be realized with uniform anisotropy field. For larger patterned ferromagnetic thin-film elements, domain formation is thought to limit the effectiveness of shape anisotropy. In our work we show that very soft lithographically patterned Ni$_{80}$Fe$_{20}$ films with control of induced magnetic anisotropy can exhibit shape anisotropy fields in agreement with single-domain models, for both hysteresis loop measurements at low field and ferromagnetic resonance measurements at high field. We show the superiority of the fluxmetric form over the magnetometric form of anisotropy estimate for thin films with control dimensions from 10 $\mu$m to 150 $\mu$m and in-plane aspect ratios above 10.

\end{abstract}

\maketitle

\indent Micron- and submicron-scale patterned ferromagnetic thin films are of interest for applications in magnetoelectronics\cite{science1,science2,patternedreview}. Shape anisotropy, through finite aspect ratios of length to width, provides a convenient mechanism to adjust the internal field $H_D=-NM$ in these structures\cite{parkin1997}, independent of induced anisotropy. In nanometer-dimension patterned structures, estimates of the demagnetizing factor from uniform (ellipsoidal) formulae\cite{jaosbornPR1945} are considered to represent the anisotropic field well\cite{carossJAPnanoellipse2011}. \\
\indent In uniformly magnetized thin-film structures of 10 $\mu$m and greater lateral dimension, appropriate formulae to estimate demagnetizing fields are not as clear. Two forms for $N$ have been proposed since the earliest treatments of the topic\cite{crmannPR1896,aharoniJAP1998,aharoniJAP2000}: "magnetometric" and "fluxmetric(ballistic)", which are the volume average and area average over the midplane normal to $\vec{M}$, respectively. These two forms diverge for structures with very large aspect ratios of thickness $b$ to in-plane dimensions $a,c$ (thin-film limit)\cite{dxchenJAP2002}. Nevertheless, the magnetometric form is widely applied in the literature\cite{carossAPL2012,zerofieldFMR}, typically without recognition that alternate forms exist. We are not aware of prior attempts to validate either form through comparison with the other in fitting experimental data.\\
\indent In this paper, we show that the fluxmetric form is in fact the superior form for the demagnetizing factor N in the thin-film limit. We have evaluated this formula both in the limit of saturated magnetization in high field, using variable frequency ferromagnetic resonance (FMR), and in low-field hysteresis, using magnetic optic Kerr effect (MOKE) measurement. We find agreement of $H_D$ to 20\% with a simple analytic limit of the formulae shown in Ref. 9. We find, on the other hand, that the magnetometric form is a very poor estimate for thin films, disagreeing with experimental data by a factor of four for the structures considered in this study.\\
\indent Patterned Py films (Ni$_{80}$Fe$_{20}$) were deposited on Si substrates using magnetron sputtering with base pressure better than $3 {\times} 10^{-9}$ Torr. The structure of the films was Si/SiO$_2$ substrate/ Ta(5 nm)/ Ni$_{80}$Fe$_{20}$(40 nm)/ Ta(3 nm). Laser-direct-write photolithography at a resolution of 1 $\mu$m was used to fabricate rectangular stripe patterns (Fig. 1a,b), with short dimensions ranging from 10 $\mu$m to 150 $\mu$m. A 1500 $\mu$m $\times$ 1500 $\mu$m square was used as the unpatterned comparison. Induced magnetic anisotropy was introduced using an \textit{in-situ} quadrupole electromagnet setup\cite{electromagnet} in UHV. A rotating field of $H_I=150$ Oe was applied in phase to the rotating sample holder at 0.25 Hz during sputtering. The Py films were deposited together on two identically patterned substrates. The first substrate, denoted as "Py-parallel" (PP), was oriented such that $H_I$ is parallel to the long axes of the elements. The second substrate, "Py-orthogonal" (PO), has the orthogonal orientation. The films were post-annealed at 250 $^{\circ}\mathrm{C}$ in vacuum of $10^{-6}$ Torr for 1 h under a field of 4.0 kOe along the deposition field to strengthen the induced anisotropy. \\
\begin{figure}[htb]
\centering
\includegraphics[width = 3.0 in] {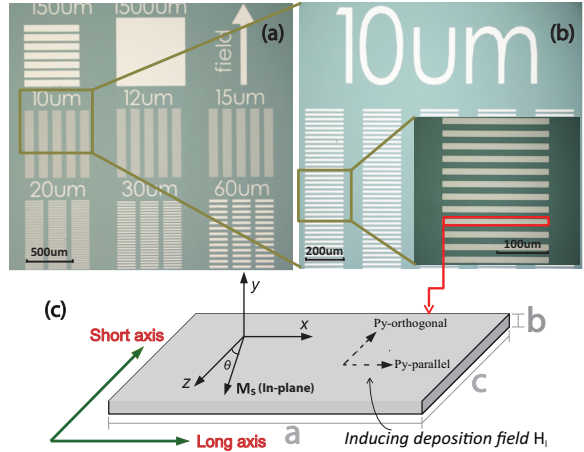}
\caption{(a,b) Patterned Ni\(_{80}\)Fe\(_{20}\) stripe arrays with short dimensions ranging from 10 $\mu$m to 150 $\mu$m. The distances between adjacent stripes are the same as their widths. (c) Definition of Cartesian coordinates and the dimensions.}
\label{fig1}
\end{figure}
\indent The hysteresis loops of the stripe arrays on PP and PO were characterized via MOKE with the biasing field applied along the short axes. The size of the laser spot is close to the array dimension (1.5 mm). The two substrates were then cut into individual arrays using a dicing saw. The in-plane shape anisotropy of the arrays was then characterized by FMR. The samples were scanned along their long and short axes. The dispersion curves $\omega(H)$ of each array were acquired with rf frequency varying from 2 Ghz to 26 Ghz. The curves were fitted by the Kittel function:
\begin{equation}
f^2=\mu_0^2\gamma^2/(2\pi)^2\cdot(H_{ex}+H_A)(H_{ex}+H_A+M_s)
\end{equation}
where $H_{ex}$ is the external biasing field, $H_A$ is the anisotropy field, $\gamma/2\pi=\textrm{2.799 Mhz/Oe}\cdot g_{eff}/2$, and $\mu_0M_s$ is the saturation inductance. The differences of the anisotropic fields ($\Delta H_A$) on the two axes are recorded in order to determine the shape anisotropy. \\
\indent To estimate $H_A$ and analyze the experimental data, consider a rectangular stripe with coordinate system defined in Fig. 1(c), with a, b, c as the long dimension, thickness, and short dimension along $\vec{x}$, $\vec{y}$ and $\vec{z}$ axes, respectively. We present a greatly simplified formula, $N_{fs}$, of the fluxmetric demagnetizing factor $N_f$ from the full analytic form (Eq. (2) in Ref. 9), which could work as a substitute for $N_f$ under thin film limit ($b<<a,c$) and when the stripe is uniformly magnetized along $\vec{z}$:
\begin{equation}
\pi{N_{fs}^z}={b\over{ac}}(\sqrt{4a^2+c^2}-c)\approx{b\over{ac}}(2a-c)
\end{equation}
The second approximation is valid when the in-plane aspect ratio is large ($a>>c$). When the magnetization is along $\vec{x}$, $N_{fs}^x$ is two orders of magnitude smaller than $N_{fs}^z$ and does not contribute to the shape anisotropy. The full analytic formulae of $N_f^z$ and $N_m^z$ from Refs. 8 and 9 are compared in Fig. 2 along with $N_{fs}^z$ for the thin film limit and a large in-plane aspect ratio (a/c=20). We highlight the significant difference between calculated magnetometric and fluxmetric forms $N_m$ and $N_f$, disagreeing with each other in the analytic calculation by a factor greater than three, as plotted in the inset of Fig. 2. We also highlight the accuracy of $N_{fs}$ to the full fluxmetric form $N_f$: the filled circles are indistinguishable from the line in Fig. 2, disagreeing with each other by a maximum of 0.03\% for the range considered in the calculation.\\
\begin{figure}[htb]
\centering
\includegraphics[width = 3.0 in] {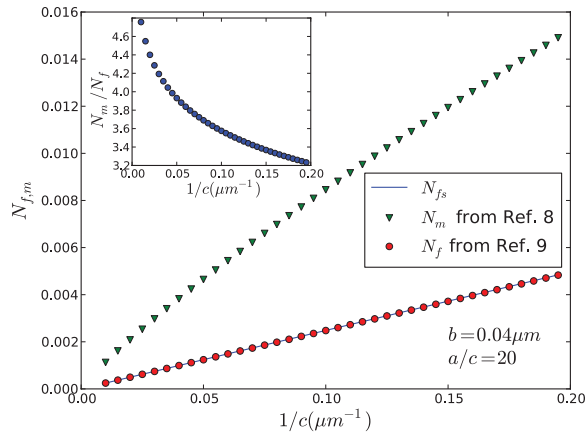}
\caption{
Comparison of $N_f$, $N_m$ and $N_{fs}$ in the thin-film limit and at large aspect ratio. Note the large divergence between $N_f$ and $N_m$ for wide samples (c $>$ 5 $\mu$m). \textit{Inset}: ratio of $N_m$ to $N_f$.}
\label{fig2}
\end{figure}
\indent In an array of thin-film elements, additional dipolar fields arise from neighboring elements. The stray dipolar field $H_s$ from the magnetostatic interaction can be calculated by summing up all the dipolar fields from the neighboring elements on each element and then taking an average of all the local fields. The stray field factor is defined as $H_s^{x,z}=-M_sN_s^{x,z}$, with the superscript denoting the magnetization direction. Parts of the calculated $\Delta N_f$ and $\Delta N_s$ are listed in Table 1, with definition of $\Delta N_{f,s}=N_{f,s}^z-N_{f,s}^x$. In the structures $\Delta N_s$ is about a quarter of $\Delta N_f$ but with opposite sign, reducing the total demagnetizing field.\\
\begin{table}[htb]
\centering
\begin{tabular}{cccccc}
\hline
array&${\Delta N_f}$&$M_s\Delta N_f$&$\Delta{N_s}$&$M_s\Delta N_s$\\
\hline\hline
"10$\mu$m"&23.4e-4&22.2Oe&-5.8e-4&-5.5Oe\\
"15$\mu$m"&15.6e-4&14.8Oe&-4.0e-4&-3.8Oe\\
"30$\mu$m"&8.5e-4&8.1Oe&-2.0e-4&-1.9Oe\\
"150$\mu$m"&1.6e-4&1.5Oe&-0.7e-4&-0.7Oe\\
\hline
\end{tabular}
\caption{Lists of calculated $\Delta N_f$ and $\Delta N_s$ for dimensions in Table 1. The equivalent anisotropic fields are calculated assuming $\mu_0M_s=0.95$ T.}
\label{table1}
\end{table}
\indent When the rectangular stripe defined in Fig. 1(c) is under an external field $H_{ex}$ along $\vec{z}$, the total energy can be expressed as:
\begin{equation}
E=-{{\mu}_0}{M_s}{H_{ex}}{\cos\theta}+[\pm K
+{{1}\over{2}}{{\mu}_0}{M_s}^2({\Delta}N^{*}+{\Delta}N_s)]{\cos^2\theta}
\end{equation}
where K is the induced anisotropy energy, $\theta$ is the angle between $\vec{M}_s$ and $\vec{H}_{ex}$, ${\Delta}N^{*}$ stands for either $\Delta N_m$ or $\Delta N_f$. The sign of the induced anisotropy energy is positive for PP and negative for PO. In a classical model, $K={\mu}_0M_sH_{k0}/2$, where $H_{k0}$ is the induced anisotropy field. The new saturation field with shape anisotropy becomes:
\begin{equation}
H_k=\pm H_{k0}+M_s(\Delta N^*+{\Delta}N_s)
\end{equation}
One should notice that when $H_k<0$ the hysteresis loop will show easy-axis behavior. \\
\indent In FMR the effective anisotropy field can be calculated by $\vec{H}_A=-{{{\partial}E_A}/{\mu_0{\partial}\vec{M}_s}}$ where $E_A$ stands for the second term in Eq. (3). The magnetization of the elements in an array will precess about $\vec{H}_{ex}$ with small amplitudes. However because the movements of different elements are not in phase, the stray dipolar fields are decoupled from the resonance precession and we approximate them as a constant field. So in Eq. (3) the term ${{1}\over{2}}{{\mu}_0}{M_s}^2{\Delta}N_s{\cos\theta^2}$ should be replaced by ${{\mu}_0}{M_s}^2{N_s^i}{\cos\theta}$, where $N_s^i$ stands for the stray dipolar factor along the biasing field direction. The effective anisotropic fields along $\vec{x}$ and $\vec{z}$ are calculated to be:
\begin{eqnarray}
H_A^{x,[z]}&={+,[-]}(\pm H_{k0}+M_s{\Delta N^*})-M_sN_s^{x,[z]}
\end{eqnarray}
At fixed frequency, the difference of the effective anisotropic field between $\vec{x}$ and $\vec{z}$ is:
\begin{equation}
{\Delta}H_A=H_A^z-H_A^x=2[\pm H_{k0}+M_s({\Delta N^*}+\Delta{N_s}/2)]
\end{equation}
\begin{figure}[htb]
\centering
\includegraphics[width = 3.0 in] {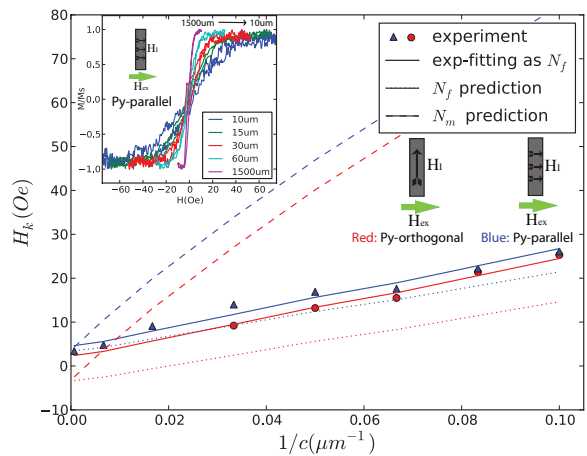}
\caption{Comparison of experimental $H_k$ from MOKE with the prediction of $N_f$ (dot) and $N_m$ (dash) from Eq. (4) for PP and PO. Solid line: data fitting into Eq. (4) by using $N_f$. \textit{Inset:} Selected MOKE loops of PP arrays.}
\label{fig3}
\end{figure}
\indent Figure 3 shows the experimental $H_k$ from MOKE, in comparison with the analytic $N_f$ (dot) and $N_m$ (dash) from Eq. (4), as a function of 1/c. The experimental $H_k$ of each array is obtained by extending the MOKE slopes at $H=0$ to saturation. For PO arrays (not shown), when the stripe width is larger than 30 $\mu$m, $H_k$ cannot be determined. The \textit{inset} shows the hysteresis loops of PO arrays, all of which show linear HA response with $H_c$ $<$ 1 Oe. The prediction curves are calculated taking the saturation inductance $\mu_0M_s=0.948$ T measured in FMR, and induced saturation field $H_{k0}=3.4$ Oe measured from the 1500 $\mu$m$\times$1500 $\mu$m square. \\
\indent The best fits to the MOKE-derived $H_k$ are shown by the solid lines. Here we treat $\mu_0M_s$ and $H_{k0}$ as free parameters but fit the two curves simultaneously. We find that the MOKE $H_k$ values are close to the $N_f$ prediction, and far from the $N_m$ prediction. The data closely approximate the fitting lines to $N_f$, validating the simple approximate form in Eq. (4). The fitted $\mu_0M_s$ is 1.17 T, 23\% larger than 0.948 T. The fitted $H_{k0}$ are 4.5 Oe(PP) and -2.3 Oe(PO). The former result is close to the 3.4 Oe but the latter is with an opposite sign. This is due to the domain wall movement rather than domain rotation and has been discovered also in patterned spin valves\cite{spinvalve}. \\
\indent Figure 4 shows the result of FMR measurement. In Fig. 4(a) the $f(H_{ex})$ relations with $\vec{H}_{ex}$ along $\vec{x}$ and $\vec{z}$ are fitted into Eq. (1) in order to extract $g_{eff}$, $H_A$ and $\mu_0M_s$. All the arrays can be fitted well by the the same $\mu_0M_s$ of 0.948 T and $g_{eff}$ of 2.115. \\
\indent The differences in $H_A$ are summarized in Fig. 4(b), in comparison with the analytic $N_f$ (dot) and $N_m$ (dash) from Eq. (7), as a function of $1/c$. The fitting into Eq. (7) is also completed with the same $\mu_0M_s$ for the two series as the fitting parameters. Very close agreement between the experimental data and fits is seen, with the fitting curves (solid lines) very close to the $N_f$ prediction (dot). The fitted $H_{k0}$ are 4.7 Oe for PP and 3.1 Oe for PO, both close to the induced saturation field 3.4 Oe. The fitted $\mu_0M_s$ is 1.08 T, 14\% larger than 0.948 T from FMR. \\
\indent {\it Summary:} we have shown that the fluxmetric demagnetizing factor $N_f$ is superior to the magnetometric form $N_m$ for large-area thin film rectangles (length/thickness $>$ 200). Furthermore, we have shown a very simple analytical approximation, $N_{fs}$, to $N_f$ in Eq. (2), which is excellent over the full range studied. These results will facilitate magnetostatic biasing of thin-film structures for applications in magnetoelectronics. \\
\begin{figure}[htb]
\includegraphics[width = 3.0 in] {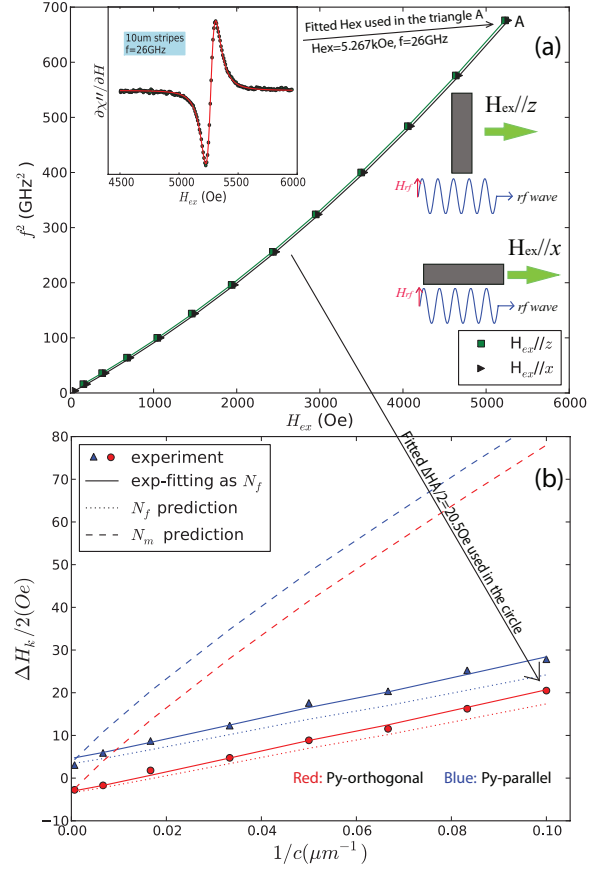}
\caption{(a) Fitting into Eq. (1) of the 10 $\mu$m PO array from in-plane FMR measurement with $H_{ex}$ along $\vec{x}$ and $\vec{z}$ axis. \textit{Inset:} Lineshape of the 10 $\mu$m PO array at 26 Ghz. (b)Comparison of experimental $\Delta H_A$ from FMR with the prediction of $N_f$ and $N_m$ from Eq. (7) for PP and PO. Same notations as in Fig. 3.}
\label{fig4}
\end{figure}
\indent We acknowledge support from the US Department of Energy grant No. DE-EE0002892 and National Science Foundation No. ECCS-0925829.

\bibliographystyle{ieeetr}



\begin{thebibliography}{5}

\bibitem{science1}
Gary A. Prinz,
Science \textbf{282}, 1660 (1998)

\bibitem{science2}
S. A. Wolf, D. D. Awschalom, R. A. Buhrman, J. M. Daughton, S. von Moln\'ar, M. L. Roukes, A. Y. Chtchelkanova, D. M. Treger,
Science \textbf{294}, 1488 (2001)

\bibitem{patternedreview}
C. A. Ross, S. Haratani, F. J. Casta\~no, Y. Hao, M. Hwang, M. Shima, J. Y. Cheng, B. Vogeli, M. Farhoud, M. Walsh, Henry I. Smith,
J. Appl. Phys. \textbf{91}, 6848 (2002)

\bibitem{parkin1997}
Yu Lu, R. A. Altman, A. Marley, S. A. Rishton, P. L. Trouilloud, Gang Xiao, W. J. Gallagher, S. S. P. Parkin,
Appl. Phys. Lett. \textbf{70}, 2610 (1997)

\bibitem{jaosbornPR1945}
J. A. Osborn,
Phys. Rev. \textbf{67}, 351 (1945)

\bibitem{carossJAPnanoellipse2011}
M. Pardavi-Horvath, B. G. Ng, F. J. Casta\~no, H. S. K{\"o}rner, C. Garcia, C. A. Ross,
J. Appl. Phys. \textbf{110}, 3921 (2011)

\bibitem{crmannPR1896}
C. Riborg Mann,
Phys. Rev. \textbf{3}, 359 (1896)

\bibitem{aharoniJAP1998}
Amikam Aharoni,
J. Appl. Phys. \textbf{83}, 3432 (1998)

\bibitem{aharoniJAP2000}
Amikam Aharoni, Ladislav Pust, Mark Kief,
J. Appl. Phys. \textbf{87}, 6564 (2000)

\bibitem{dxchenJAP2002}
D. X. Chen, C. Prados, E. Pardo, A. Sanchez, A. Hernando,
J. Appl. Phys. \textbf{91}, 5254 (2002)

\bibitem{carossAPL2012}
G. Shimon, A. O. Adeyeye, C. A. Ross,
Appl. Phys. Lett. \textbf{101}, 083112 (2012)

\bibitem{zerofieldFMR}
Y. Zhuang, M. Vrouble, B. Rejaei, J. N. Burghartz,
J. Appl. Phys. \textbf{99}, 08C705 (2006)

\bibitem{electromagnet}
C. Cheng, N. Sturcken, K. Shepard, W. E. Bailey,
Rev. Sci. Instrum. \textbf{83}, 3903 (2012)

\bibitem{spinvalve}
Z. H. Qian, R. Bai, C. M. Yang, Q. L. Li, Y. C. Sun, D. X. Huo, L. W. Li, H. L. Zhan, Y. Li, J. G. Zhu,
J. Appl. Phys. \textbf{109}, 103904 (2011)

\end{thebibliography}
\end{document}